\documentclass{appolb}
\usepackage{graphicx}

\begin{document}
\title{Fission Fragments Mass Yields of Actinide Nuclei
\thanks{Presented at the Zakopane Conference on Nuclear Physics, Zakopane 2022.}
\author{Krzysztof Pomorski\footnote{Email: Krzysztof.Pomorski@umcs.pl, 
ORCID ID: 0000-0002-5557-6037}, Bo\.zena Pomorska}
\address{Institute of Physics, Maria Curie Sk{\l}odowska University, 
            20-031 Lublin, Poland}}
\maketitle
\begin{abstract}
A new, rapidly convergent Fourier over spheroid parametrization is developed to describe the shape of a fissioning nucleus: its elongation, non-axiality and left-right asymmetry and neck formation. The 4D Potential Energy Surfaces (PES) of even-even actinide nuclei are evaluated within the macro-micro model. The Langevin trajectories generated on such PESs allow for obtaining the fission fragments' mass and kinetic energy yields. The charge equilibration at the scission configuration and the post-scission neutron emission are also discussed.
\end{abstract}
  
\section{Introduction}

The nuclear fission phenomenon, discovered in 1938, is still very attractive to experimentalists and theoreticians. Accurate reproduction of such observables as the mass and the total kinetic energy yields of fission fragments or the multiplicities of emitted neutrons is a good test of any modern theoretical model. A review of the existing fission models can be found, e.g., in Refs. \cite{MJV19,RMo13,SJA16,BBB20,MSc17,ACD21}. Readers interested in the theory of nuclear fission can find more details in the textbook \cite{KPo12}.

The present research is a continuation of our previous works \cite{PIN17,PDH20,PBK21,LCW21,KDN21} in which the fission fragment mass-yields of nuclei from different mass regions were obtained using the Fourier shape parametrization \cite{SPN17}. In the present paper, we extend our 3D Langevin model by adding a mode responsible for the charge equilibration of the fragments and a Master type equation describing the multiplicity of emitted neutrons. In addition, an innovatory, better adopted to make the fission calculation on a grid, the Fourier-over-Spheroid (FoS) shape parametrization (see also \cite{PNB17}) is used. One has to stress here that the nuclear shapes generated by FoS are equivalent to those given by the Fourier parameterization \cite{SPN17}. The main features of our model are described in Section 2, while the neutron emission from the fragments is discussed in Section 3, followed by the Summary.

\section{Nuclear shape parametrization}

The potential energy surfaces of fissioning nuclei are obtained using the macro-micro model. The macroscopic part of the energy is evaluated according to the Lublin-Strasbourg-Drop (LSD) formula \cite{PDu09}, while the microscopic energy corrections are calculated using the Yukawa-folded single-particle potential \cite{DPB16}.

The surface of the fissioning nucleus is described in the cylindrical coordinates $(\rho,\varphi,z)$ by the following formula:  
\begin{equation}
\rho^2(z,\varphi)=\frac{R_0^2}{c}\,f\left(\frac{z-z_{\rm sh}}{z_0}\right)
{1-\eta^2\over 1+\eta^2+2\eta\cos(2\varphi)} ~,
\label{rhos}
\end{equation}
where $\rho(z,\varphi)$ is the distance from the $z$-axis to the surface.
Function $f(u)$ defines the shape of the nucleus having half-length $c=1$:
\begin{equation}
\begin{array}{rl}
  f(u)&=1-u^2-\left({a_4\over 3}-{a_6\over 5}+\dots\right)
   \cos\left({\pi\over 2}u\right)-a_3\sin(\pi\,u) \\[2ex]
    &-a_4\cos\left({3\pi\over 2}u\right)
     -a_5\sin(2\pi\,u)-a_6\cos\left({5\pi\over 2}u\right)-\dots~,
\end{array}
\label{fos}
\end{equation}
where $-1\leq u \leq 1$. The first two terms in $f(u)$ describe a sphere, the third ensures volume conservation for arbitrary deformation parameters $\{a_3,\;a_4,\;\dots\}$. The parameter $c$ determines the elongation of the nucleus keeping its volume fixed, while $a_3$ and $a_4$ describe the reflectional asymmetry and the neck size, respectively. The half-length is $z_0=cR_0$, where $R_0$ is the radius of a sphere with the same volume. The $z$-coordinate varies in the range $-z_0+z_{\rm sh}\leq z\leq z_0+ z_{\rm sh}$. The shift $z_{\rm sh} = -3/(4\pi) z_0 (a_3-a_5/2+\dots)$ places the nuclear center of mass at the origin of the coordinate system. The parameter $\eta$ describes a possible elliptical, non-axial deformation of a nucleus. 

The formula (\ref{rhos}) is completely equivalent to those based on the Fourier expansion and described in Refs.~\cite{SPN17}. Here, the deviation from a sphere with radius $\rho=1$ is firstly expanded in the Fourier series, and subsequently this deformed object of the length $2R_0$ is scaled to the elongation equal to $2cR_0$. The formula (\ref{rhos}) is more adapted to calculation of the PES made on a mesh in the multidimensional deformation parameter ($c,a_3,a_4,...,a_n$) space, since the range of variability of the $a_k$ coefficients does not depend on the elongation $c$. In addition, the mass ratio of the fragments, their relative distance, and the radius of the neck between them, measured in $z_0$ units, do not depend on the elongation of the nucleus. It is also worth noticing that for the reflection symmetric shapes, the geometrical scission points appear when $a_4=a_4^{\rm sc}={3\over 4}+{6\over 5}a_6\dots$ independently of the elongation $c$. Such properties of the present FoS shape parametrization make it very useful for all kinds of calculations related to nuclear fission.

\begin{figure}[htb]
\includegraphics[width=12.5cm]{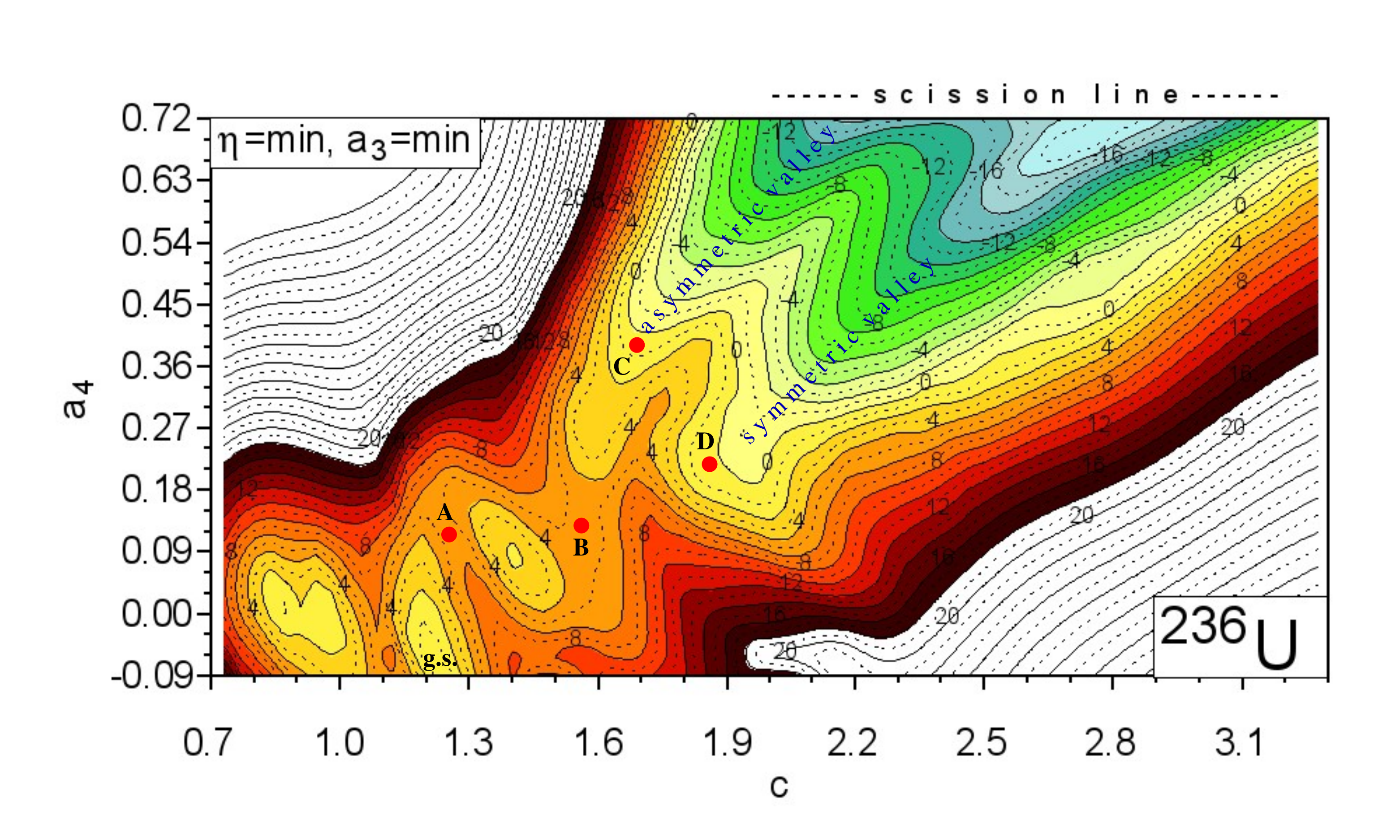}\\[-4ex]
\caption{Potential energy surface of $^{236}$U on the ($c,\,a_4$) plane. Each point is minimized with respect to the non-axial ($\eta$) and the reflectional ($a_3$) deformations.}
\label{Fig1}
\end{figure}

A typical PES of a fissioning nucleus is shown in Fig.~\ref{Fig1}. It is a projection of the 4D PES onto the $(c,a_4)$ plane, i.e., each energy point in the $(c,a_4)$ map of $^{236}$U is minimized with respect the non-axial $\eta$ and reflectional $a_3$ deformation parameters. The ground state (g.s.), first (A), and second (B) saddle points, as well as the exit points from the fission barrier leading to the asymmetric (C) and symmetric (D) fission valleys, are marked. The upper value of the neck-parameter $a_4=0.72$ corresponds to the neck radius approximately equal to the nucleon radius. The non-axial degree of freedom is important at a smaller elongation of the nucleus until the neighborhood of the second saddle. At larger deformation, their effect is negligible, which allows us to restrict the Langevin calculations to 3D when discussing the dynamics of fission. Moreover, the role of the higher-order deformation parameters $a_5$ and $a_6$ is rather small even in the region of well-separated fission fragments as it was shown in Ref.~\cite{KDN21}.\\[-5ex]

%
%
\subsection{Mass and TKE yields obtained within 3D Langevin calculation}

The Langevin equation governs the dissipative fission dynamics. In the generalized coordinates ($\{q_i\},~~i=1,2,...,n$) it has the following
form \cite{KPo12}:
\begin{equation}
\begin{array}{l}
 {dq_i\over dt} = \sum\limits_{j} \; [{\cal M}^{-1}(\vec q\,)]_{i\, j} \; p_j  \\
 {dp_i\over dt} =  - {1\over 2} \sum\limits_{j,k} \,
           {\partial[{\cal M}^{-1}]_{jk}\over\partial q_i}\; p_j \; p_k
          -{\partial V(\vec q)\over\partial q_i}
          - \sum\limits_{j,k} \gamma_{ij}(\vec q) \;
           [{\cal M}^{-1}]_{jk} \; p_k + F_i(t) \,\,, 
\end{array}
\label{LGV}
\end{equation}
Here $V(\vec q\,)=E_{\rm pot}(\vec q\,)-a(\vec q\,)T^2$ is the free-energy of
fissioning nucleus having temperature $T$ and the single-particle level density
$a(\vec q\,)$. In the present calculation the inertia ${\cal M}_{jk}$ and the friction $\gamma_{ij}$ tensors are evaluated in the irrotational flow and the wall approximation, respectively, as described in Refs.~\cite{BNP19,KDN21}.
\begin{figure}[htb]
\includegraphics[width=0.5\textwidth]{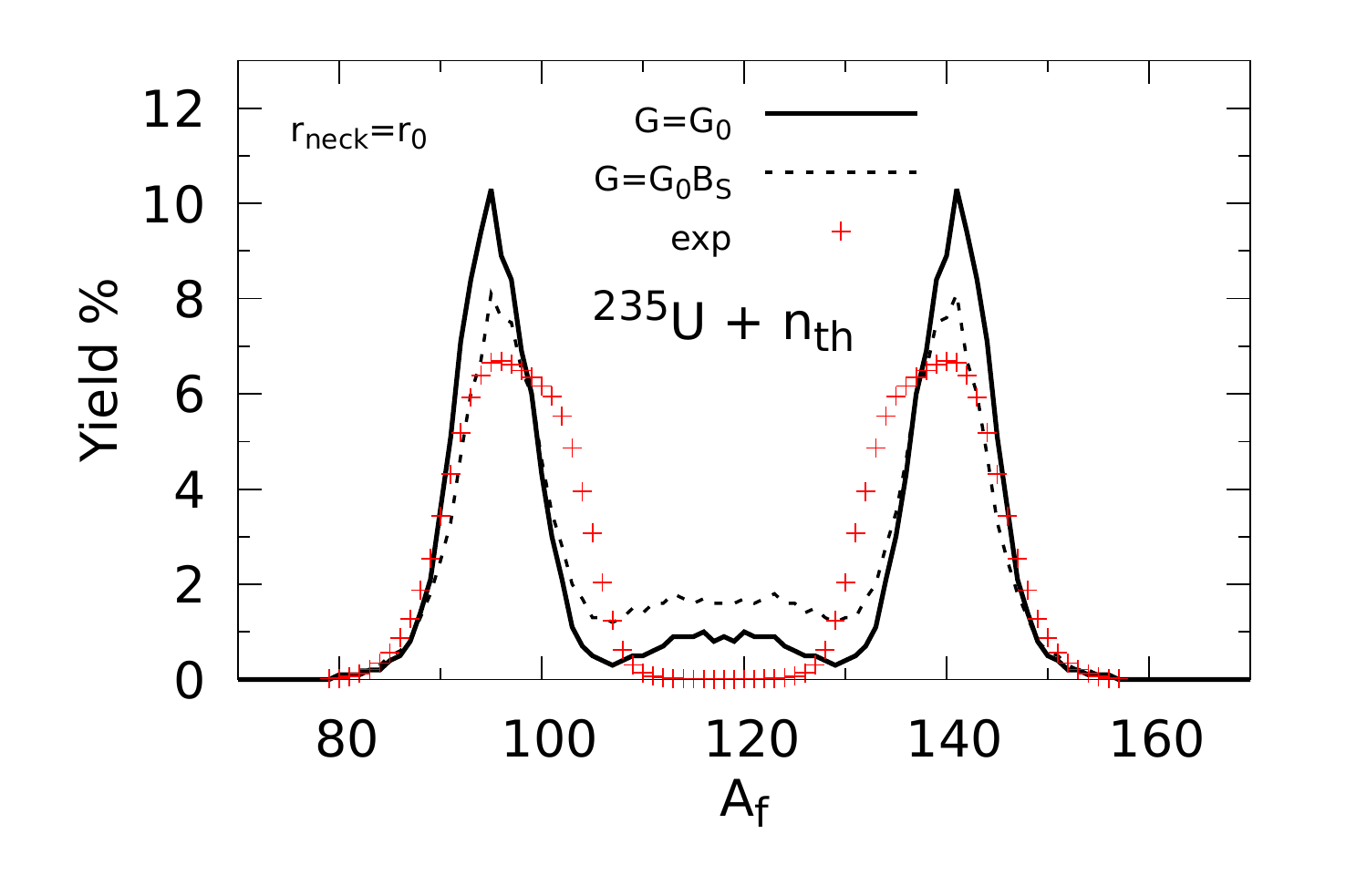}
\includegraphics[width=0.5\textwidth]{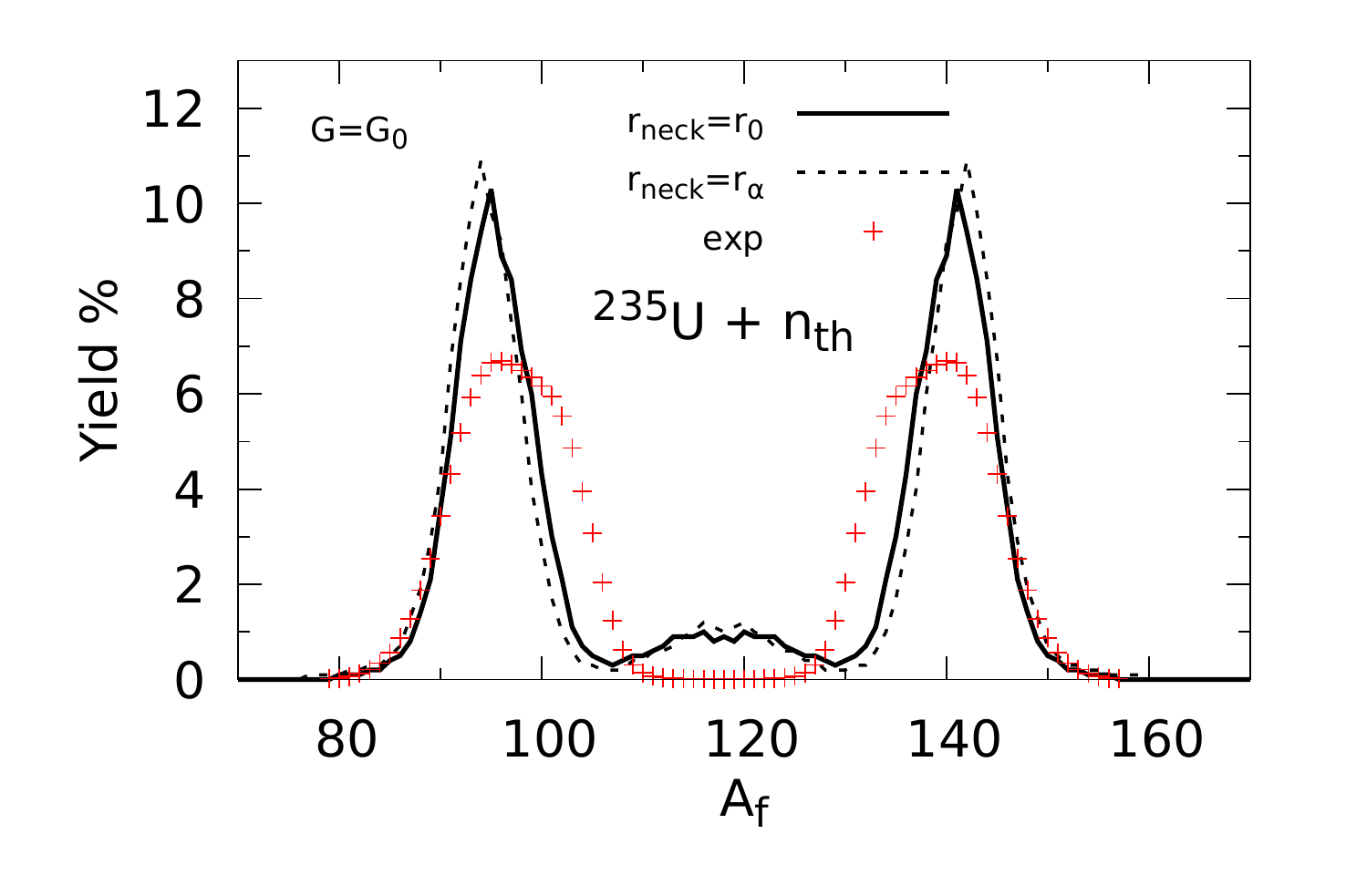}\\[-5ex]
\caption{Fission fragment mass-yield of $^{235}$U + n$_{\rm th}$ as a function 
of the mass of the fragment. The l.h.s. figure shows the yields obtained assuming the constant (solid line) and the proportional to the surface of the nucleus (dashed line) pairing strength. Similar results obtained assuming two different neck radii (radius of nucleon or alpha-particle) at which the fission occurs are shown in the r.h.s. panel. The experimental data are taken from Ref. \cite{DKT69}.}
\label{Fig2}
\end{figure}

The vector $\vec F(t)$ stands for the random Langevin force, which couples the collective dynamics to the intrinsic degrees of freedom and is defined as:
\begin{equation}
F_i(t) \!\!=\!\! \sum_{j} g_{ij}(\vec q\,) \; G_j(t) \,\,,
\label{rforce}
\end{equation}
where $\vec G(t)$ is a stochastic function whose strength
$g(\vec q\,)$ is given by the diffusion tensor ${\cal D}(\vec q\,)$ 
defined by the generalized Einstein relation:
\begin{equation}
{\cal D}_{ij} \!\!=\!\!T^*\gamma_{ij} \!\!=\!\! \sum_{k} g_{ik} \; g_{jk}~,~~~
{\rm where}~~~T^*=E_0/{\rm tanh}\left({E_0\over T}\right)~.
\label{Eirel}
\end{equation}
Here $E_0$= 1 MeV is the zero-point collective energy. The temperature $T$ is obtained from the thermal excitation energy ($E^*$) equal to the difference between the initial ($E_{\rm init}$ ) and the total collective energy, beeing the sum of the kinetic ($E_{\rm kin}$) and potential ($V$) energies of the fissioning nucleus at a given deformation point ($\vec q$):\\[-2ex] 
\begin{equation}
 a(\vec q\,)T^2=E^*(\vec q\,)=E_{\rm init}-(E_{\rm kin}+V)~.
\label{temp}
\end{equation}
Running thousands of random Langevin trajectories, which end at the scission configuration, one can estimate the distribution of the mass and total kinetic energies of the fragments. An example of our estimates of the fission fragment mass yields obtained in the thermal neutron-induced fission of $^{235}$U is shown in Fig.~\ref{Fig2}.\\[-4ex]
\begin{figure}[htb]
\begin{center}
\includegraphics[width=0.5\textwidth]{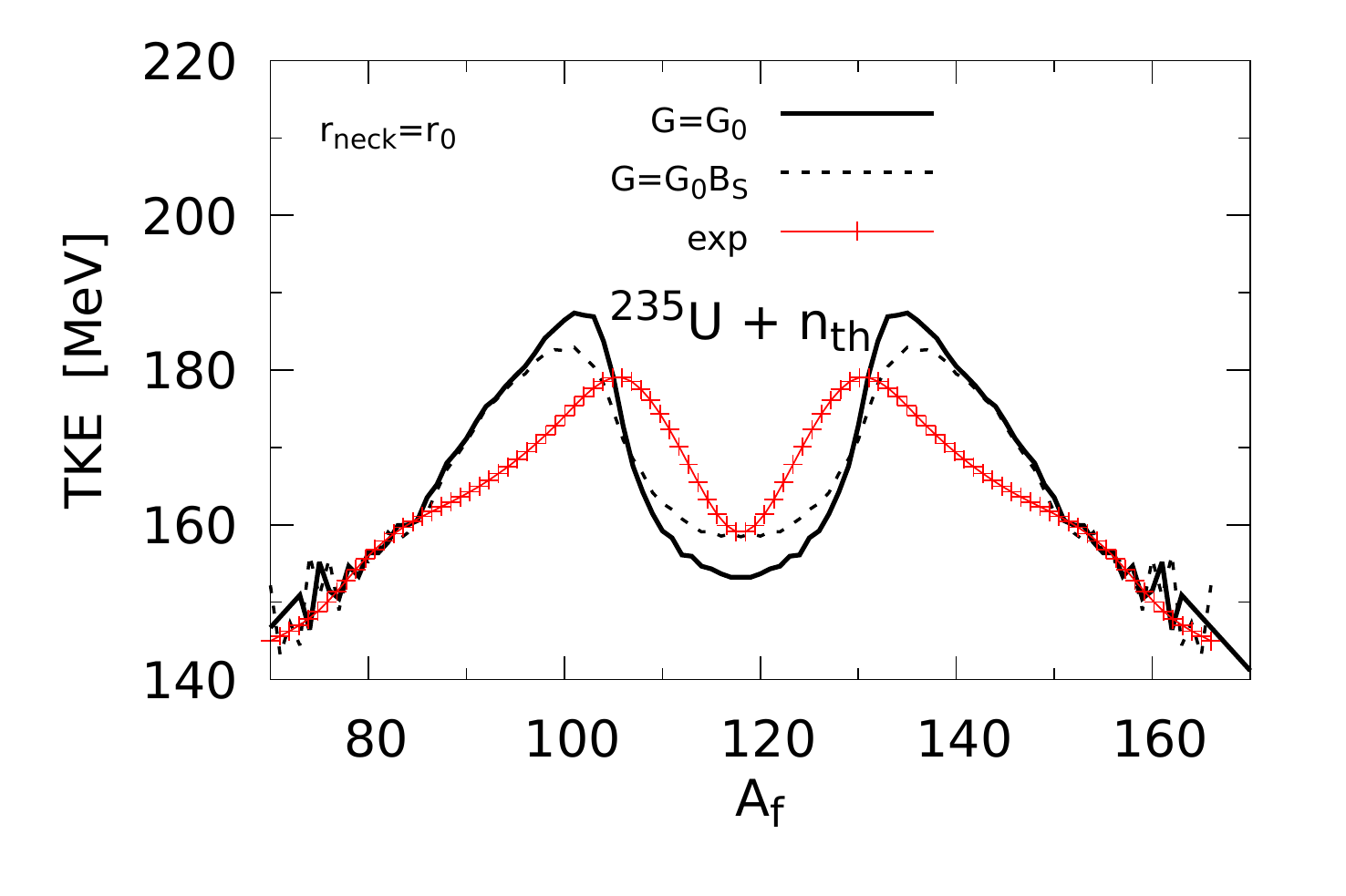}\\[-3ex]
\caption{Total kinetic energy of fission fragments of $^{235}$U + n$_{\rm th}$ as a function of the fragment mass. The solid and the dashed lines correspond to the constant and the proportional to the surface of nucleus pairing strength, respectively. The experimental data (+) are taken from Ref. \cite{ATJ20}.}
\end{center}
\label{Fig3}
\end{figure}

The fission fragments' total kinetic energy (TKE), shown in Fig. 3, is approximately given by their Coulomb repulsion energy. This energy is equal to the difference between total Coulomb energy of the nucleus at the scission configuration and Coulomb energies of both deformed fragments: 
\begin{equation}
{\rm TKE}={3e^2\over 5r_0}\left[{Z^2\over A^{1/3}}B_{\rm Coul}(\vec q_{\rm sc})
           -{Z_{\rm h}^2\over A_{\rm h}^{1/3}}\,B_{\rm Coul}(\vec q_{\rm h})
           -{Z_{\rm l}^2\over A_{\rm l}^{1/3}}\,B_{\rm Coul}(\vec q_{\rm l}) \right]~.
\label{Ekin}
\end{equation}  
It is undoubtedly a more accurate estimate of the fission-fragment kinetic energy than the frequently used its point-charge approximation: $E_{\rm kin}=e^2 Z_{\rm h}Z_{\rm l}/R_{12}$, where $R_{12}$ is the distance between the fragment mass-centers.


\subsection{On the charge equilibration at scission}

Knowing the fragment deformation at scission, it is possible to find a
preferred charge for each fragment. In the majority of Langevin-type calculations, one assumes that the ration $N/Z$ of the fragments is the same as the one of the fissioning nucleus. Looking at the proton and neutron microscopic density distributions, one could obtain better estimates, which is rather hard to do. Below, we propose a simple model for the proton-neutron equilibrium at scission based on the LD and the pairing correlation energy. Such charge equilibration can be determined by looking at the change of the total energy of the fissioning system with the charge number of the heavy fragment $Z_{\rm h}$:
\begin{equation}
\begin{array}{rl} 
E(Z,A,Z_{\rm h};A_{\rm h},\vec q_{\rm h},\vec q_{\rm l})
    &=E_{\rm LD}(Z_{\rm h},A_{\rm h};\vec q_{\rm h})
     +\,E_{\rm LD}(Z-Z_{\rm h},A-A_{\rm h});\vec q_{\rm l})\nonumber\\[+1ex]
    &+\,e^2Z_{\rm h}(Z-Z_{\rm h})/R_{12}-E_{\rm LD}(Z,A;0)~,
\end{array}
\label{echeq}
\end{equation}
where $Z, A$, and $Z_{\rm h}, A_{\rm h}$ are the charge and mass numbers of the mother nucleus and the heavy fragment, respectively. The mass as well as the deformation parameters of the heavy ($A_{\rm h},\,\vec q_{\rm h}$) and the light fragments ($A_{\rm l},\,\vec q_{\rm l}$) are fixed by the shape of the nucleus at scission corresponding to the end of each Langevin trajectory.

\begin{figure}
\includegraphics[width=0.5\textwidth]{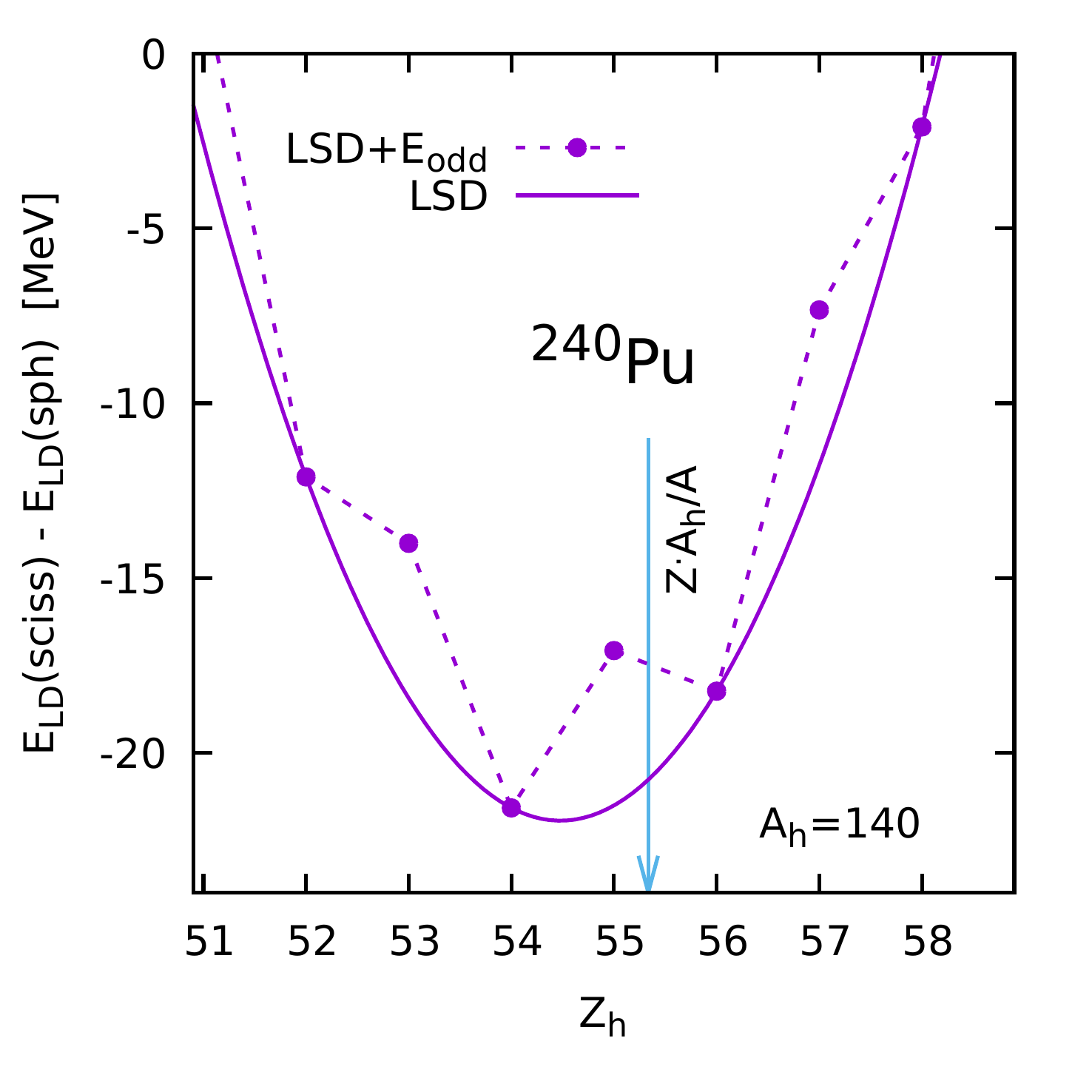}\hfill
\includegraphics[width=0.5\textwidth]{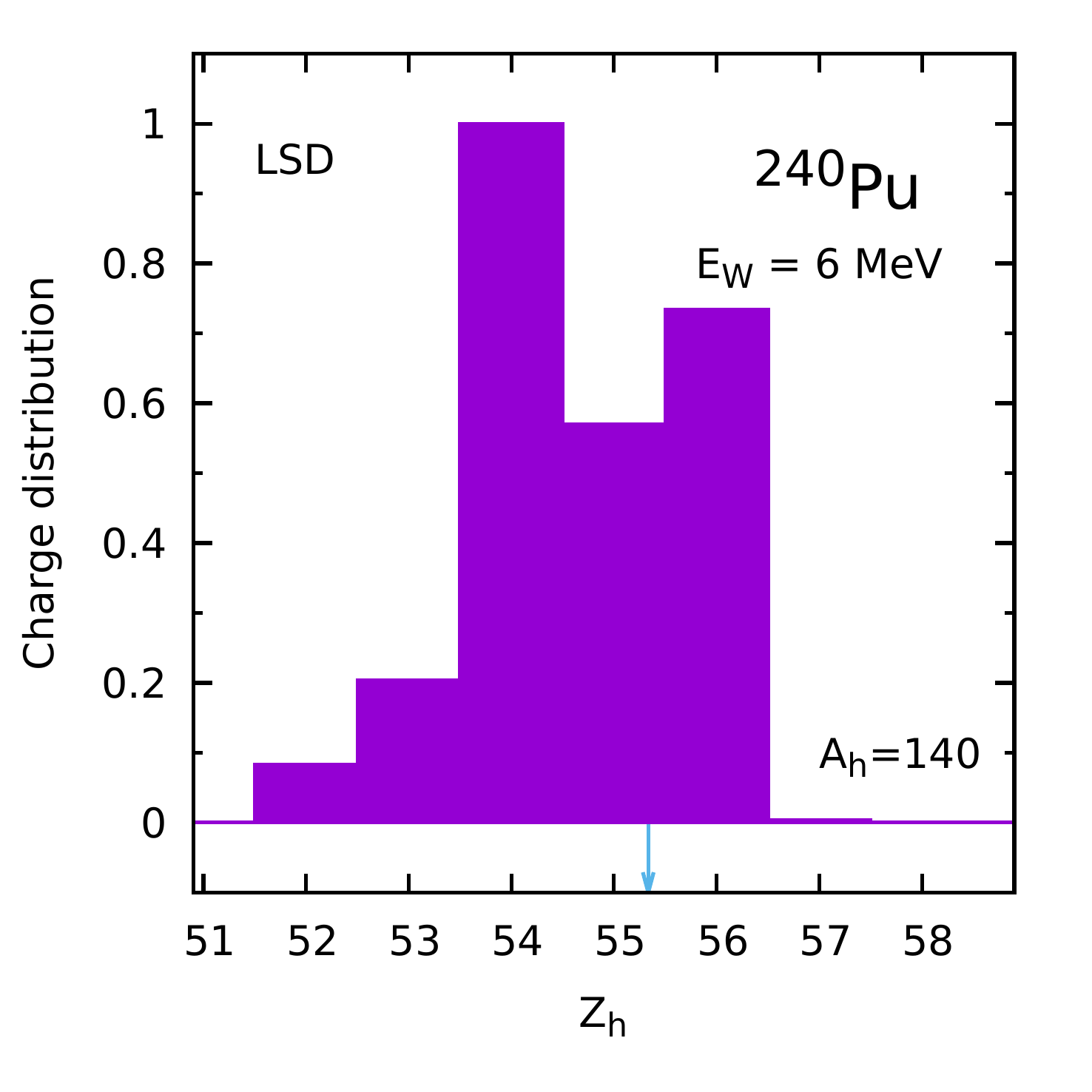}\\[-4ex]
\caption{Energy of $^{240}$Pu at scission as a function of the heavy fragment charge number in the LSD mass formula \cite{PDu09} (l.h.s.) and the Wigner distribution probability of the fragment charge number (r.h.s.).}
\label{Fig4}
\end{figure}

The total energy as a function of the fragment charge number is shown in the l.h.s. of Fig.~\ref{Fig4}.
The distribution of the heavy-fragment charge number can be estimated using a Wigner function corresponding to the energy $E$ given by Eq.~\ref{echeq} for different
values of $Z_{\rm h}$:
\begin{equation}
 W(Z_{\rm h})=\exp\{-[E(Z_{\rm h})-E_{\rm min}]^2/E_{\rm W}^2]~,
\label{Wigner}
\end{equation}
which gives the distribution probability of the fragment charge shown in
the r.h.s. of Fig.~\ref{Fig4}. $E_{\rm min}$ in Eq.~\ref{Wigner} is the lowest discrete energy as a function of $Z_{\rm h}$. Furthermore, the following random number decides about the charge number $Z_{\rm h}$ of the heavy fragment, with $Z_{\rm l}=Z-Z_{\rm h}$. The energy $E_{\rm W}$ should be comparable with the energy distance $\hbar\omega_0$ between harmonic oscillator shells since we have here a single-particle (proton) transfer between the touching fragments.

\begin{figure}[htb]
\includegraphics[width=0.5\textwidth]{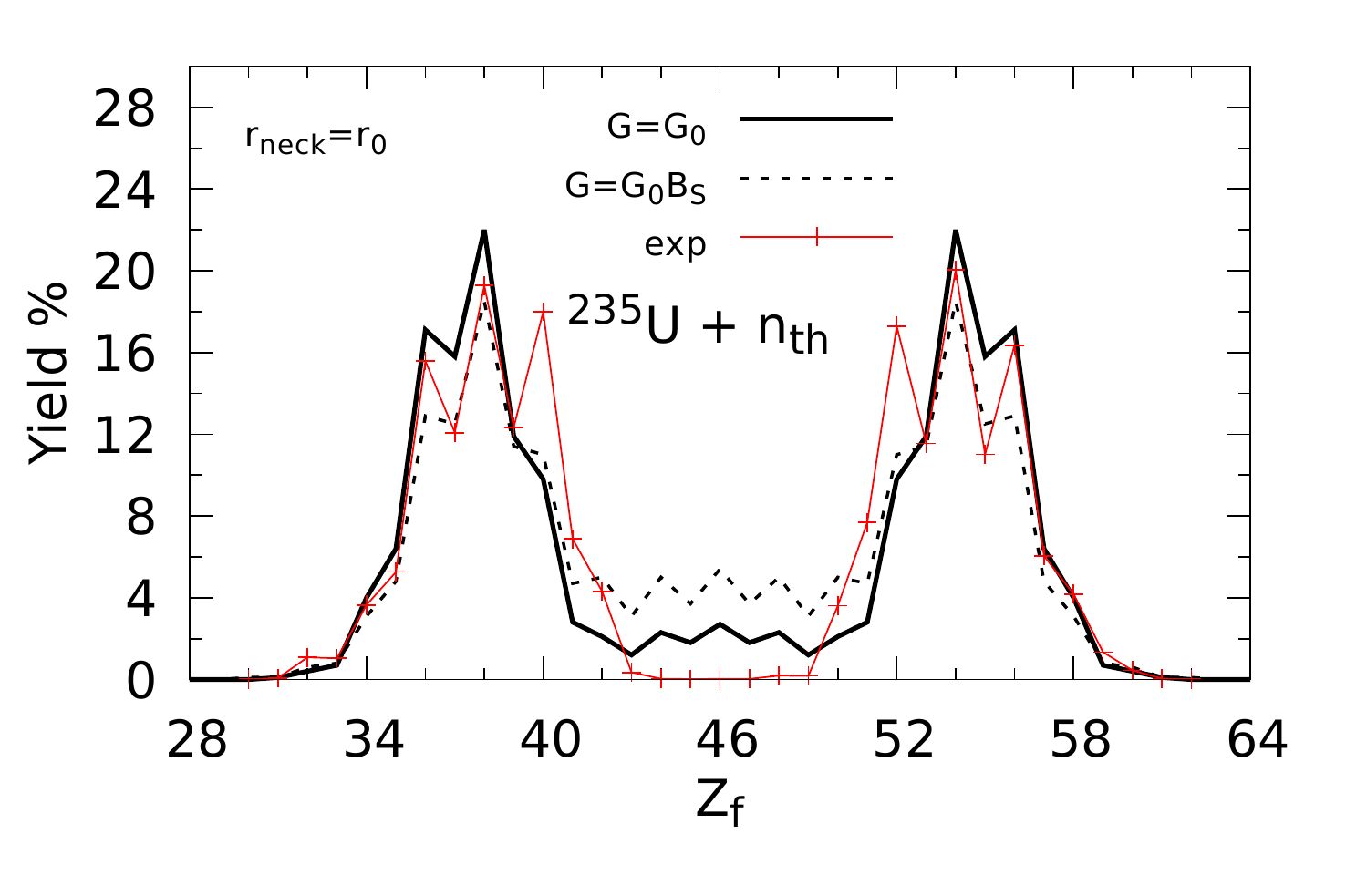}
\includegraphics[width=0.5\textwidth]{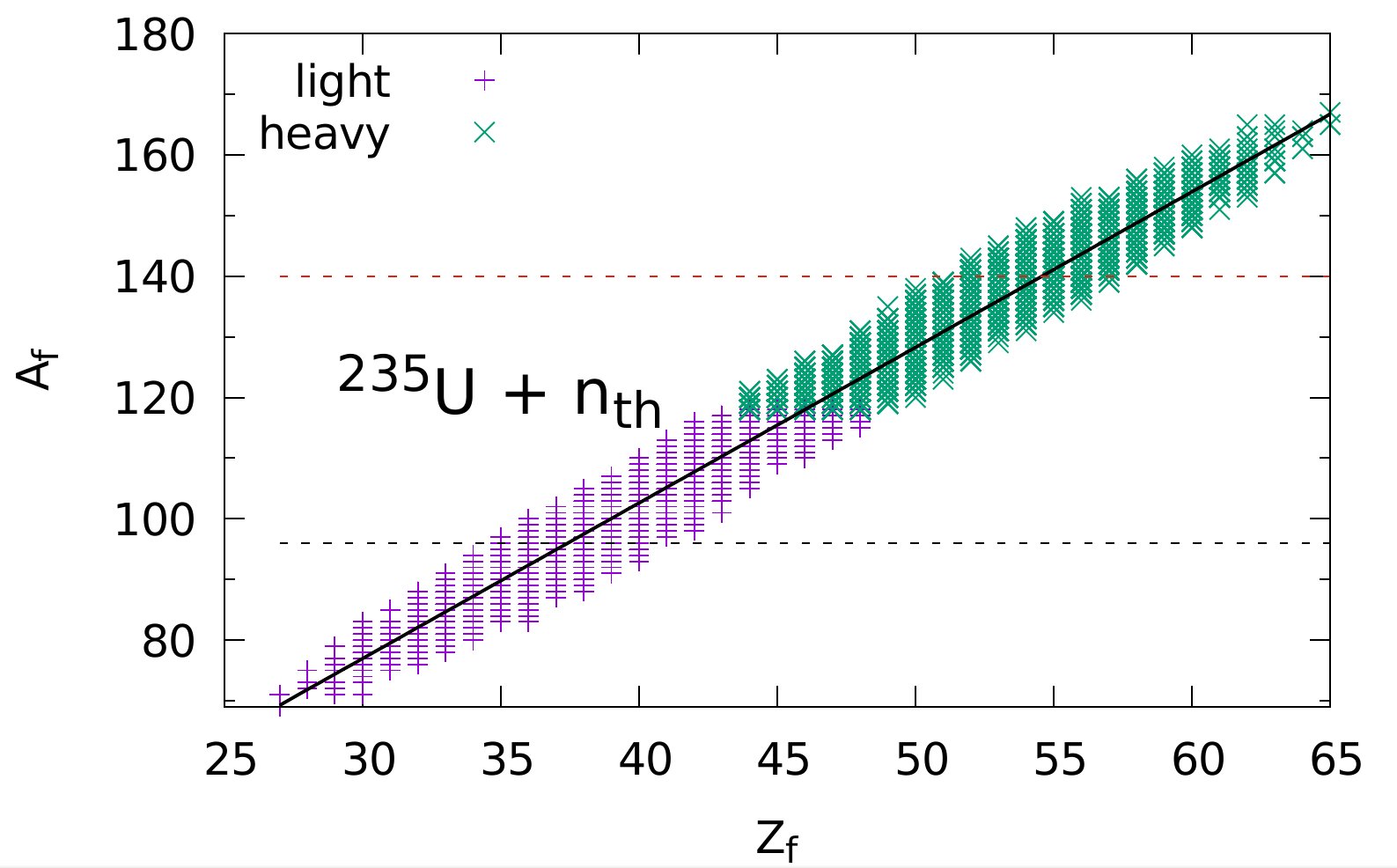}\\[-4ex]
\caption{The fission fragment yield of $^{235}$U + n$_{\rm th}$ (l.h.s.) as a function of the fragment charge. The solid and the dashed lines correspond respectively to the constant and the proportional to the surface of nucleus pairing strength. The experimental data (+) are taken from Ref. \cite{MSc17}. The correlation between the fragment mass and charge is displayd in the r.h.s. figure.}
\label{Fig5}
\end{figure}

The above charge equilibration effect has to be considered at the end of each Langevin trajectory when one fixes the integer fragment mass and charge numbers of the fission fragments. The resulting fission fragment yield is compared with the data \cite{MSc17} in the l.h.s. part of Fig.~\ref{Fig5}, while the correlation between fragment mass and charge is displayed in the in the r.h.s. figure.

\begin{figure}[htb]
\begin{center}
\includegraphics[width=0.6\textwidth]{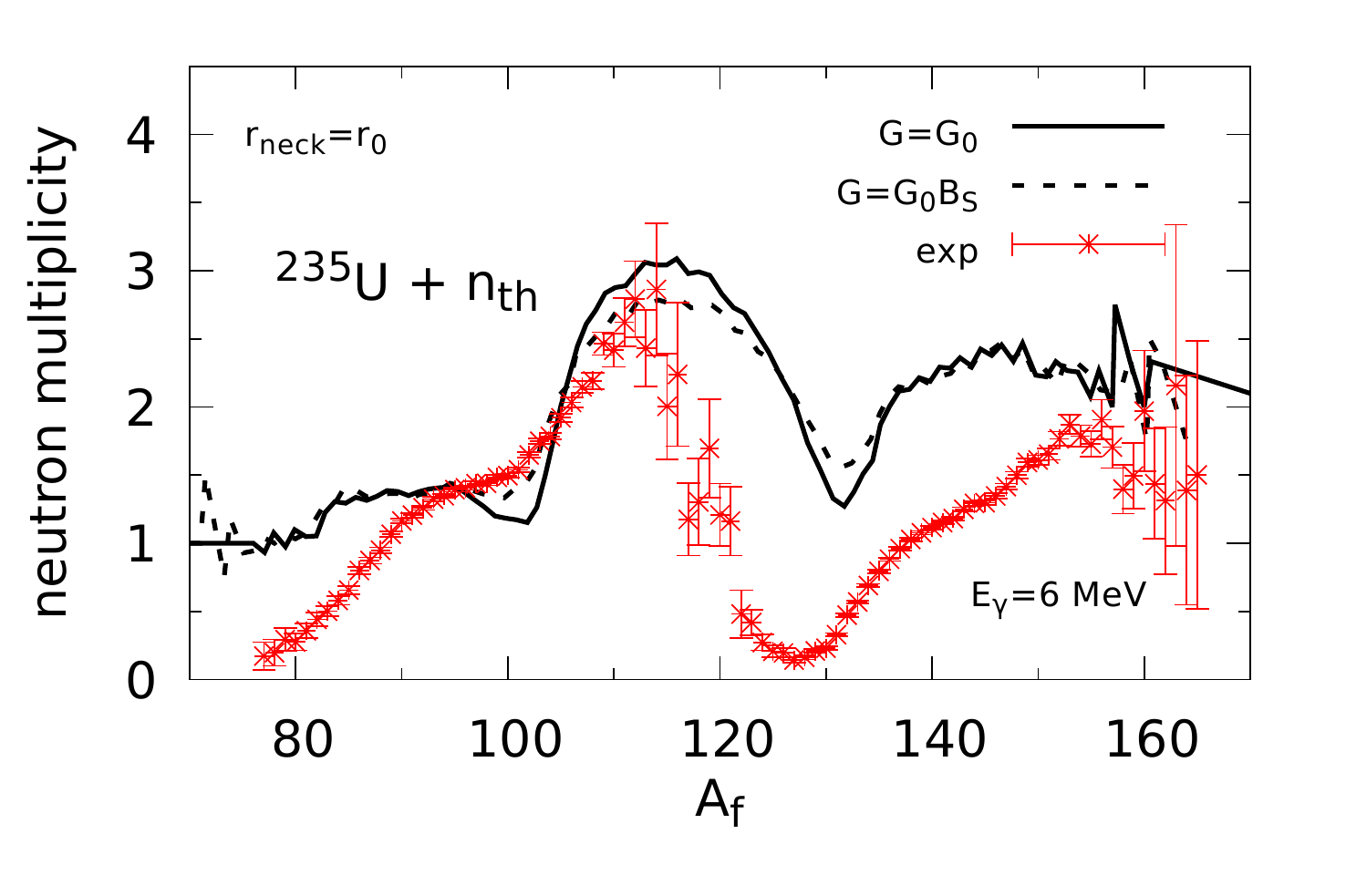}\\[-8ex]
\end{center}
\caption{Multiplicity of neutrons emitted by the fission fragments of $^{235}$U + n$_{\rm th}$ as a function of the mass of the fragment. The solid and the dashed lines correspond respectively to the constant and the proportional to the surface of nucleus pairing strength. The experimental data (*) are taken from Ref.~\cite{GHO18}.}
\label{Fig6}
\end{figure}


\section{Neutron emission from the fission fragments}

The maximal energy of a neutron emitted from a fragment (mother) can be 
obtained from the energy conservation law:
\begin{equation}
\epsilon_{\rm n}^{\rm max}= M_{\rm M}+E_{\rm M}^*-M_{\rm D}-M_{\rm n}~,
\label{En}
\end{equation}
where $M_{\rm M},\,M_{\rm D},\,M_{\rm n}$ are the mass excesses of mother and daughter nuclei, and of the neutron, respectively. These data can
be taken from a mass table \cite{KWH21}.
The thermal excitation energy of the  daughter nucleus is: 
\begin{equation}
 E_{\rm D}^*=\epsilon_{\rm n}^{\rm max}-\epsilon_{\rm n}~.
\label{Eexc}
\end{equation}
Here $e_{\rm n}$ is the kinetic energy of emited neutron.

One assumes that the thermal energy of a fragment $E^*_i$ in the scission point is proportional to its single-particle level density:
\begin{equation}
 {E^*_{\rm l}\over E^*_{\rm h}}= {a(Z_{\rm l},A_{\rm l};{\rm def}_{\rm l})\over a(Z_{\rm h},A_{\rm h};{\rm def}_{\rm h})}~~~~
{\rm and}~~E^*=a(Z,A;{\rm def})\,T^2=E^*_{\rm l}+E^*_{\rm h}~.
\label{Eteq}
\end{equation}
The deformation energy of each fragment can be evaluated in the LD model:
\begin{equation}
E_{\rm def}^{(i)}\approx E_{\rm LD}(Z_i,A_i,{\rm def}_i)
                 - E_{\rm exp}(Z_i,A_i,{\rm g.s.})~.
\label{Edefi}
\end{equation}
The total excitation energy ($E^{(i)}_{\rm exc}$) of fragment $i$ is then the sum of  its thermal and its deformation energy:
\begin{equation}
 E_{\rm exc}^{(i)}=E_{\rm def}^{(i)}+E_i^*~.
\label{Eexci}
\end{equation}
This energy is converted into heat due to the presence of the friction
force, which allows to evaluate the effective temperature $T_i$ of each 
fragment:
\begin{equation}
 E_{\rm exc}^{(i)}=a(i)\,T_i^2~,
\label{Ti}
\end{equation}
where $a(i)$ is the single-particle level density of the $i$-th fragment.
These data enable us to estimate the number of neutrons emitted from each fragment.

The neutron emission probability is given by the Wei{\ss}kopf formula \cite{Del86}:
\begin{equation}
\Gamma_{\rm n}(\epsilon_{\rm n})={2\mu\over\pi^2\hbar^2\rho_{\rm M}
(E^*_{\rm M})}\int\limits_0^{\epsilon_{\rm n}}\sigma_{\rm inv}(\epsilon)\,
\epsilon\,\rho_{\rm D}(E^*_{\rm D})\,d\epsilon~.
\label{Wei}
\end{equation}
Here $\mu$ is the reduced mass of the neutron, $\sigma_{\rm inv}$ is the
neutron inverse cross-section \cite{DFF80}:
\begin{equation}
\sigma_{\rm inv}(\epsilon)=[0.76+1.93/A^{1/3} + (1.66/A^{2/3}-0.050)/\epsilon]\,\pi\,(1.70A^{1/3})^2~,
\label{ncross}
\end{equation}
while  $\rho_{\rm M}$ and $\rho_{\rm D}$ are, respectively, the level densities of mother and daughter nuclei:
\begin{equation}
\rho(E)={\sqrt{\pi}\over 12a^{1/4}E^{5/4}}\exp(2\sqrt{aE})~,
\label{lden}
\end{equation}
Here the single-particle level density parameters $a$ of the mother and the daughter are taken here from Ref.~\cite{Ner02}. 

Our estimates of the neutron multiplicity obtained for $^{235}$U + n$_{\rm th}$
is compared in Fig.~\ref{Fig6} with the experimental data \cite{GHO18}.


\section{Summary}

We have shown that our 3D model based on the Fourier-over-Spheroid shape parametrization describes reasonably the main features of the low-energy fission of atomic nuclei. Typical results for the thermal neutron induced fission of $^{235}$U shown above illustrate well the quality of our approach. The following conclusions can be drawn from our investigation:\\[-4ex]
\begin{itemize}
\item Fourier expansion of nuclear shape offers a very effective way 
      of describing the shapes of fissioning nuclei both in the vicinity of the 
      ground-state as well as in the scission point.\\[-4ex]
\item The potential energy surfaces are evaluated in the macro-micro model using
      the LSD formula for the macroscopic part of the energy and the 
      Yukawa-folded single-particle to obtain the microscopic energy 
      correction.\\[-4ex] 
\item It was shown that a 3D Langevin model, which couples the fission, 
      neck and mass asymmetry modes, describes the main features 
      of the fragment mass and kinetic energy yields.\\[-4ex]
\item The distribution of the multiplicity of neutrons emitted by the fragments 
      are reproduced well in our model.\\[-2ex] 
\end{itemize}
Further calculations for wider mass and excitation energy ranges of fissioning nuclei are in progress.\\[3ex]

\parindent 0pt
{\bf Acknowledgments}\\[1ex]
We acknowledge discussions with F. A. Ivanyuk and C. Schmitt, and thank A. G\"o\"ok and A. Al-Adili for providing experimental data. This work was supported by the National Science Centre, project No. 2018/30/Q/ST2/00185.


\end{document}